\documentclass[twocolumn,aps,prc,superscriptaddress,showpacs]{revtex4}
\usepackage{amsmath,bm}
\usepackage{graphicx}

\begin{document}

\draft
%\preprint{}
\title{Exploring  binding energy and separation energy dependences of HBT strength }

\author{ Y. B. Wei}\thanks{Email: weiyb@sinr.ac.cn.}
\affiliation{Shanghai Institute of Applied Physics, Chinese
Academy of Sciences, P.O. Box 800-204, Shanghai 201800, China}
\author{Y. G. Ma} \thanks{Corresponding Author.   Email: ygma@sinr.ac.cn}
  \affiliation{Shanghai Institute of Applied Physics, Chinese Academy of Sciences, P.O. Box 800-204,
Shanghai 201800, China}
   \affiliation{   CCAST (World Laboratory), PO Box 8730, Beijing 100080, China}
\author{ W. Q. Shen}
  \affiliation{Shanghai Institute of Applied Physics, Chinese Academy of Sciences, P.O. Box 800-204,
Shanghai 201800, China}
\affiliation{   CCAST (World Laboratory), PO Box 8730, Beijing 100080, China}
  \affiliation{  Department of Physics, Ningbo University, Ningbo 315211, China}
\author{G. L. Ma}
\author{K. Wang}
\author{X. Z. Cai}
\author{ C. Zhong}
\author{ W. Guo}
\author{J. G. Chen}
\affiliation{Shanghai Institute of Applied Physics, Chinese
Academy of Sciences, P.O. Box 800-204, Shanghai 201800, China}

 \date{\today}

\begin{abstract}
Hanbury Brown-Twiss (HBT) results of the nucleon-nucleon
correlation function have been  presented for the nuclear
reactions with neutron-rich projectiles (Be isotopes) using an
event-generator, the Isospin-Dependent Quantum Molecular Dynamics
model. We explore that the relationship between the binding energy
per nucleon of the projectiles and the strength of the
neutron-proton HBT at small relative momentum. Moreover, we reveal
the relationship between the single neutron separation energy and
the strength of the halo neutron-proton HBT. Results show that
neutron-proton HBT results are sensitive to binding energy or
separation energy.
\end{abstract}
　　
 \maketitle

\section{Introduction}

Hanbury Brown-Twiss (HBT) technique was presented for the
astrophysical measurements a few decades ago, and it reveals
information about the angular diameters of distant stars
\cite{Brown}. More recently, it has been widely used in others
fields, for instance,  the analogous correlations in
semiconductors and in free space aiming at the fermionic
statistics of electrons \cite{Henny,Picciotto,Oliver}. It has also
become an important tool in high energy region since it can be
utilized to measure the evolving geometry of the interaction zone
while being applied to the studies to search for a possible
quark-gluon plasma and study the properties of the predicted new
state of matter \cite{Heinz1}. In the past several years it has
also significant theoretical development and widespread
application in subatomic physics
\cite{Goldhaber,Koonin,Pratt,Sullivan,Achim,Boal,Heinz2}. The
emission time and source size in the nuclear reaction can be
extracted by the nuclear HBT technique, which is one kind of
intensity interferometry. In the applications of experimental and
theoretical heavy ion reactions at intermediate energy,  various
aspects have been investigated via the correlation functions, such
as the dependences  of the isospin of the emitting source
\cite{Ghetti}, the impact parameter \cite{Gong},  the nuclear
symmetrical energy \cite{Chen1} and the total momentum of nucleon
pairs \cite{Colonna} and so on. The details about the EOS and the
collision process can be revealed from the correlation functions
well.

Some groups have applied HBT technique to study the exotic nuclear
reaction recently \cite{Orr,Marques1,Marques2}. Studies have been
performed for many years for exotic nuclei with the increasing
availability of radioactive nuclear beams around the world. Among
the various techniques to investigate on the exotic nuclei, the
measurements of the total interaction cross section and the
fragment momentum distribution of the projectile are the main
methods to explore the exotic structure in the past years
\cite{Pochodzalla,Tanihata1,Tanihata2,Kobayashi,Suzuki,Tanihata_sum,Zahar,Shen,Ma1,Ma2,Arnell}.
Some nuclei like $^{6}He$ \cite{Tanihata1,Kobayashi}, $^{11}Li$
\cite{Tanihata1}, $^{14}Be$
\cite{Suzuki,Tanihata_sum,Zahar,Marques2} are considered as
two-neutron halo ones and $^{11}Be$ \cite{Tanihata2,Kobayashi},
$^{15}C$ \cite{Ozawa1,Bazin}, $^{22}O$ \cite{Ozawa2} are as one
neutron halo, and the proton halo structure has been also  proved
to exist in the structure of the nuclei $^{8}B$ \cite{Tanihata1},
$^{12}N$ \cite{Warner}, $^{23}Al$ \cite{Cai} etc. In terms of the
structure of the halo, integral measurements, such as total
reaction cross sections, are only sensitive to the overall size.
Dissociation reactions, in which the core and/or nucleons are
detected in the final state, can provide some structure
information \cite{Orr}-the major difficulty being the relationship
between the initial and final states as dictated by the distorting
effects of the reaction \cite{Marques1}. So it is very interesting
to investigate the exotic nuclei via HBT technique further. As we
know, the binding energy and the nucleon(s)-separation energy are
important for the structure of the nuclei. The former indicates
the stable level of the nucleus and the nucleon-nucleon
relationship among the nucleus, and the later is a good criterion
to verify the possibility of the exotic nucleus. These two have
been studied through the density calculation by RMF theory in the
past years. Researches about these two factors via the nuclear
collisions systematics are needed. In this Letter, we expect to
explore the relationship between these two factors and the
nucleon-nucleon correlation function value at very small relative
momentum with help of Isospin-Dependent Quantum Molecular Dynamics
(IDQMD)  model which can describe the reaction dynamics on event
by event basis.

\section{HBT Technique}

Firstly, we would like to  recall the HBT technique. 　As we know
the wave function of relative motion of light identical particles,
when emitted in close proximity in space-time, is modified by the
final-state interaction (FSI) and quantum statistical symmetries
(QSS), and this is the principle of the intensity interferometry,
i.e. HBT. The correlation function is defined as the ratio between
the measured two-particle distribution and the product of the
independent single-particle distributions:
 \begin{equation}
              C(p_1,p_2)  = \frac{dn^2/dp_1dp_2}{dn/dp_1 dn/dp_2},
\end{equation}
where $d^2n/dp_1dp_2$ represents the correlated two-particle distribution and
$dn/dp_1$ and $dn/dp_2$ is the independent single-particle
distributions of particle 1 and 2, respectively.
 Usually, the projection $C(q) = c_0 \frac{N(q)}{D(q)}$ onto the relative three-momentum
($q = \frac{1}{2}|\vec{p_1}-\vec{p_2}|$) is used as the momentum
correlation function, where the measured distribution of pairs (
N(q) ) is divided by a reconstructed distribution of uncorrelated
pairs ( D(q) ). $c_0$ is a normalized coefficient so that $C(q)$
tends to 1 at high relative momentum, where the effects of FSI and
QSS vanish. The deviation of $C(q)$ from 1 thus reflects the
information of the emission source. Other effects, arising from
the form of the single-particle distributions or the experimental
acceptances, are eliminated by the denominator of Eq. (1). 　

　Interpretation of correlation functions measured in heavy-ion
collisions requires understanding the relationship of the
parameters extracted from fitting the data and the true
single-particle distributions at freeze-out. This relationship can
be established by using an event generator that models the
collision dynamics, particle production, and then constructing a
two-particle correlation function. The event-generator correlation
functions are constructed from the positions and momenta
representing the single-particle emission distribution at the time
of the last strong interaction, i.e. at freeze-out. The
event-generator used in our task is the Isospin-dependent Quantum
Molecular Dynamics transport model, which has been applied
successfully to the studies of the heavy-ion collisions which
includes the simulation of the reaction process
\cite{Aichelin,Peilert,Li,Ma3,Zhang,Liu,Singh}. For completeness,
we  would like to make a brief introduction to this model in
following section.

\section{Model Description}

The Quantum Molecular Dynamics (QMD) approach  is an n-body theory
to describe heavy ion reactions from intermediate energy to 2
GeV/n. It includes five important parts: the initialization of the
target and the projectile; the propagation in the effective
potential; the collisions between the nucleons; the Pauli blocking
effect and the numerical tests. A general review about the QMD
model can be found in \cite{Aichelin}.
 The IDQMD model is based on the
QMD model affiliating the isospin factors.

As we know, the dynamics in heavy-ion collisions (HIC) at
intermediate energies is mainly governed by three components, the
mean field, two-body collisions, and pauli blocking. Therefore,
for an isospin-dependent reaction dynamics model it is important
for these three components to include isospin degrees of freedom.
What is more essential, in initialization of projectile and target
nuclei, the samples of neutrons and protons in phase space should
be treated separately because there exists a large difference
between neutron and proton density distributions for nuclei far
from the $\beta$-stability line. Particularly, for neutron-rich
nucleus one should sample a stable initialized nucleus with
neutron-skin structure and therefore one can directly explore the
nuclear structure effects through a microscopic transport model.
The IDQMD model has been improved based on the above ideas.

    In the IDQMD model, the nuclear mean field can be parametrized by
\begin{equation}
U(\rho,\tau_{z}) = \alpha(\frac{\rho}{\rho_{0}}) +
\beta(\frac{\rho}{\rho_{0}})^{\gamma} +
\frac{1}{2}(1-\tau_{z})V_{c} \nonumber
\end{equation}
\begin{equation}
+ C\frac{(\rho_{n} - \rho_{p})}{\rho_{0}}\tau_{z} + U^{Yuk}
\end{equation}
with $\rho_{0}$ the normal nuclear matter density (here, 0.16
$fm^{-3})$, $\alpha$ = -124 MeV, $\beta$ = 70.5 MeV and $\gamma$ =
2.0. $\rho$, $\rho_{n}$ and $\rho_{p}$ are the total, neutron, and
proton densities, respectively; The coefficients $\alpha$ and
$\beta$ are parameters for nuclear equation of states (EOS).
$\tau_{z}$ is the $z$th component of the isospin degree of
freedom, which equals 1 or -1 for neutrons or protons,
respectively; $V_{c}$ is the coulomb potential; and $U^{Yuk}$ is
Yukawa (surface) potential which has the following form:

\begin{equation}
U^{Yuk}=\frac{V_{y}}{2m}\sum_{{i}\neq{j}}\frac{1}{r_{ij}}exp(Lm^{2})[exp(-mr_{ij})erf(\sqrt{L}m
\nonumber
\end{equation}
\begin{equation}
-r_{ij}/\sqrt{4L})-exp(mr_{ij})erf(\sqrt{L}m+r_{ij}/\sqrt{4L})]
\end{equation}
with $V_{y}$ = 0.0074 GeV, m = 1.25 $fm^{-1}$ and L = 2.0 $
fm^{2}$. The relative distance $r_{ij} = |\vec{r_i}-\vec{r_j}|$.
In our task, the so-called soft EOS with an incompressibility of K
= 200 MeV is used and the symmetry strength C = 32 MeV
\cite{Aichelin}.

The NN cross section is the experimental parametrization which is
isospin dependent. Recently, studies of collective flow in HIC at
intermediate energies have revealed the reduction of the in-medium
NN cross sections \cite{Westfall,Klakow}. An empirical
expression of the in-medium NN cross section \cite{Klakow} is
used:
\begin{equation}
\sigma_{NN}^{med} = (1 + f \frac{\rho}{\rho_{0}}) \sigma_{NN}^{free}
\end{equation}
with the factor $f$ $\approx$ - 0.2 which has been found to better
reproduce the flow data \cite{Westfall}. Here $\sigma_{NN}^{free}$
is the experimental NN cross section \cite{Chen3}. The
neutron-proton cross section is about three times larger than the
neutron-neutron or proton-proton cross section below 300
MeV/nucleon.

The Pauli blocking effect taken in IDQMD model is treated
separately according to the neutron and the proton: Whenever a
collision has occurred, in the phase space we assume that each
nucleon occupies a six-dimensional sphere with a volume of
$\hbar^{3}$/2 (considering the spin degree of freedom), and then
calculate the phase volume, V, of the scattered nucleons being
occupied by the rest nucleons with the same isospin as that of the
scattered ones. We then compare 2V/$\hbar^{3}$ with a random
number and decide whether the collision is blocked or not.

For the initialization of the nucleons of the target and
projectile, the IDQMD model distinguishes the proton and neutron
from each other. The neutron and the proton density distribution
are determined from the Skyrme-Hartree-Fock (SHF) method with
parameter set $SKM^{*}$ which can give reasonable density
distribution for stable and neutron-rich nuclei \cite{Wen}. One
can obtain the radial positions of nucleons in the initial nuclei
in terms of the Monte-Carlo method. In the model, the radial
density can be written as:
 \begin{equation}
\rho(r) = \sum_{i}\frac{1}{(2\pi
L)^{3/2}} exp(-\frac{r^{2} + r_{i}^{2}}{2L})\frac{L}{2rr_{i}}
\nonumber
\end{equation}

 \begin{equation}
\times[exp(\frac{rr_{i}}{L}) - exp(-\frac{rr_{i}}{L})]  .
\end{equation}

 And the
momentum distribution of nucleons is generated by means of the
local Fermi gas approximation. The local Fermi momentum   is given
by:
 \begin{equation}
P^{i}_{F}(\vec r) = \hbar(3\pi^{2}\rho_{i}(\vec r))^{\frac{1}{3}},
(i= n,p).
\end{equation}

    The stability of the propagation of the initialized nuclei has
been checked in details and can last at least 200 fm/c according
to the evolutions of the average binding energies and the root
mean square radii of the initialized nuclei.

\section{Results and Discussions}

    Among the exotic nuclei, $^{14}$Be nucleus, with four protons and
ten neutrons has received the most attentions, both theoretically
and experimentally, due to its rather unique structure
\cite{Suzuki,Tanihata_sum,Zahar,Marques2}. To test our approach,
firstly we have analyzed the reaction of $^{14}$Be fragments into
$^{12}$Be + 2n at 35 MeV/nucleon with the target $^{12}C$. Fig. 1
shows the experimental and the calculated correlation functions
for neutron and neutron. The two halo neutrons in calculation are
defined as the emitted neutrons in coincidence with $^{12}$Be
core. The solid line is the calculated two halo neutron
correlation function.  It shows clearly that the correlation
function between the two halo neutrons reproduces the experimental
data excellently, which is consistent with the small two neutron
separation energy. This directly leads to the neutron density
distribution of $^{14}$Be extending very large which is a major
reason for the abnormal larger total reaction cross section and
the very narrow momentum distribution of the fragment $^{12}$Be
\cite{Tanihata_sum}. In our work, the nucleons are defined as
emitted if they do not belong to any clusters ($A \geq 2$) which
 are recognized  by a simple coalescence model: ie. nucleons are
considered to be part of a cluster if in the end at least one
other nucleon is closer than $r_{min} \leq 3.5$ fm in coordinate
space and $p_{min} \leq 300$ MeV/c  in momentum space in the final
state \cite{Aichelin,Liu}. This definition of nucleon emission is
crucial for the good reproduction of the experimental data in Fig.
1. The $^{14}Be$ is therefore a good test case for the HBT method.

\begin{figure}
\vspace{-0.6truein}
\includegraphics[scale=0.4]{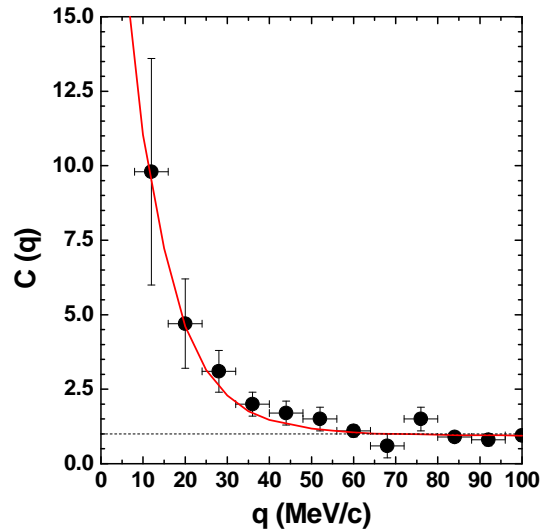}
\vspace{-0.4truein} \caption{\footnotesize The solid circles with
error bars are the two halo-neutron correlation functions (hh) in
the collision of $^{14}Be$ fragmented into $^{12}Be$ + 2n at 35
MeV/n  and the target is $^{12}C$\cite{Marques2}. The solid line
is the calculated two-halo neutrons correlation function.}
\label{Fig1}
\end{figure}

Based upon the achievement of halo neutron - halo neutron
correlation function of  HBT results in IDQMD with the data, we
further explore the proton-neutron correlation function as a
function of binding energy or separation energy. The target is
$^{12}C$ and the projectile is Be isotope. Only those events in
which the neutron and the proton are emitted in the same event are
accepted. The calculated results are shown in Fig. 2. The figure
shows the proton-neutron correlation function for different Be
isotopes and the insert of Fig. 2 shows the relationship between
the strength of proton-neutron correlation function $C_{PN}$ at 5
MeV/c and the binding energy per nucleon of the projectile
$E_{binding}$. The solid line of the insert is just a linear fit
to guide the eyes.

    One can notice that the behavior of the correlation functions
between the proton and the neutron as a function of neutron number
and binding energy in Fig. 2. Generally the strength of HBT at
very small relative momentum shows a clear dependence on the
neutron number. The tendency of the $C_{PN}$ rises with the
increasing $E_{binding}$. As we know, among the projectiles we
studied, the number of the protons is 4 and that of the neutron
are gradually increasing, and this will reflect the stability of
the nuclei. With the increasing of the mass number, the mean
relationship between the nucleons will change weaker. Since the
strength $C_{PN}$ of correlation function symbolizes the mean
relationship between the emitted proton and neutron and the
binding energy per nucleon associated with the tightness between
the nucleon, the tendency shown in the insert of Fig. 2 reflects
that the $C_{PN}$ can reveal the compactness of the nuclei.

\begin{figure}
\vspace{-0.6truein}
\includegraphics[scale=0.4]{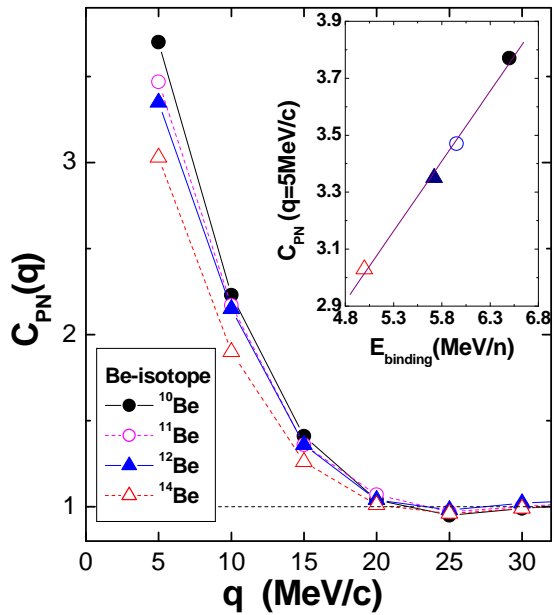}
\vspace{-0.2truein} \caption{\footnotesize The proton-neutron
correlation function for different Be isotopes. The meaning of
symbols are illustrated in left bottom corner of Figure. The
insert shows the relationship between the strength of
proton-neutron correlation function $C_{PN}$ at 5 MeV/c and the
binding energy per nucleon of the projectile $E_{binding}$. The
collisions were simulated at 800 MeV/n of the incident energy and
head-on collisions. The target is $^{12}$C.} \label{Fig2}
\end{figure}

On the other hand, we also study the correlation functions between
the proton and the most outside neutron. Here the most outside
neutron is defined as the one which is the most far away from the
spatial center at the FSI and it is also called as a halo neutron
for simplification, even though it is not strict in physics sense.
To obtain reasonable HBT results for halo neutron and proton, only
those events in which the halo neutron and proton are emitted in
the same event are accepted to investigate such a correlation
function. The similar correlation function was obtained as
$C_{PN}$ and the relationship between the strength of proton-halo
neutron correlation function $C_{PH}$ at 5 MeV/c and the
single-neutron separation energy of the projectile $E_{sep}$ was
extracted. The symbols of Fig. 3 show the calculated  result and
the solid line is just a linear fit to guide the eyes. It looks
that, with the increasing of the $E_{sep}$, $C_{PH}$ at 5 MeV/c
rises gradually. Since the strength $C_{PH}$ of correlation
function between neutron and proton increases with the decreasing
of the source size, the above $E_{sep}$ dependence of $C_{PH}$
reflects that the emission source size of the neutron and proton
shrinks with the separation energy of the single neutron, which is
consistent with the extent of binding of single neutron via
$E_{sep}$.

\begin{figure}
\vspace{-0.8truein}
\includegraphics[scale=0.4]{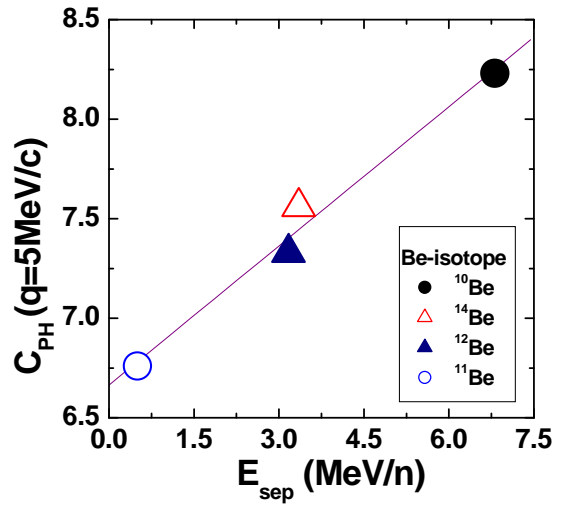}
\vspace{-0.8truein} \caption{\footnotesize
The relationship between the proton-halo neutron correlation
function $C_{PH}$ at 5 MeV/c and the single-neutron
separation energy of the nuclei. The
solid line is just a linear fit.
The collisions were simulated at 800
MeV/n of the incident energy and head-on collisions.
The target is $^{12}$C.  }
\label{Fig3}
\end{figure}

　 One should keep in mind that the $E_{sep}$ is thought as one of
the important information for halo structure, for Be isotope, the
nucleus $^{11}Be$ is considered to exist the halo configuration
according to its extremely small one-neutron separation energy.
Compared with other nuclei in Be isotopes, the $C_{PH}$ at small
relative momentum of it is smaller, i.e. there exists a larger
spatial diffusion. That indicates that HBT of proton and halo neutron
can show the different configuration of the nuclei Be isotopes
well. These observations might illustrate that the $C_{PH}$ will symbolize the
relationship between the most outside neutron and the emitted proton,
 which is related to the extension level of
the mass density distribution directly. Unfortunately, in the
practical point of view, this correlation function $C_{PH}$ is
almost impossible to be measured. However, it still bear some
information in theoretical point of view.

\section{Summary}

In summary, the intensity interferometry (HBT) technique has been
applied to investigate its sensitivity to the binding energy and
separation energy of neutron-rich nuclei   from the break-up of
nuclei by convoluting the phase-space distribution generated with
the IDQMD model. Firstly we gave a well-fitted halo neutron - halo
neutron correlation function from the break-up of $^{14}Be$ on $C$
target. Based upon this achievement of the good fit, we explore
the dependence of the proton-neutron correlation function
($C_{PN}$) at small relative momentum with the binding energy
($E_{binding}$) for Be isotopes. It was found that the correlation
strength of $C_{PN}$ at small relative momentum rises with the the
binding energy. This changeable tendency of $C_{PN}$ with
$E_{binding}$ is here reported for the first time and it might be
a potential good way to study the structure of the nuclei.
Moreover, the proton-halo neutron correlation function ($C_{PH}$)
is also constructed from the break-up reactions. There exists the
similar relationship between the $C_{PH}$ at small relative
momentum and the separation energy ($E_{sep}$) of Be isotopes as
the relationship of $C_{PN}$ to $E_{binding}$. From theoretical
point of view, $C_{PH}$ at small relative momentum is sensitive to
$E_{sep}$ and this can be attributed to the spatial extension
level of the neutron which is most far away the center of the
nucleus. Of course, we recognize that it is very difficult to
measure $C_{PH}$ in  a practical point. However, theoretical
behavior is also interesting.

\begin{acknowledgments}
This work was supported in part
 by the Major State Basic Research Development Program under
 Contract No G200077400, the Chinese Academy of Sciences Grant for the
Distinguished Young Scholars of National Natural Science
Foundation of China under Grant No 19725521, the National Natural
Science Foundation of China under Grant No 10135030 and the
Shanghai Phosphor Program Under Contract Number 03 QA 14066. Y. G.
Ma would like to appreciate Dr. Scott Pratt for providing CRAB
code which is used to construct the momentum correlation function
from phase space data.
\end{acknowledgments}

{}

\end{document}